\documentclass[runningheads]{llncs}

\usepackage{todonotes}
\usepackage[T1]{fontenc}
\usepackage{graphicx}
\usepackage{hyperref}
\usepackage{color}

\usepackage{listings}
\usepackage{caption}
\usepackage{subcaption}
\usepackage{booktabs}

\begin{document}
\title{Oh SSH-it, what's my fingerprint?\\ A Large-Scale Analysis of SSH Host Key Fingerprint Verification Records in the DNS}
\titlerunning{A Large-Scale Analysis of SSHFP records}

\author{Sebastian Neef\inst{1}\orcidID{0000-0003-3055-0823}
	\and Nils Wisiol\inst{1}\orcidID{0000-0003-2606-614X}}
%
\authorrunning{S. Neef et al.}
%
\institute{Security in Telecommunications, Technische Universität Berlin, Germany
\email{neef@tu-berlin.de}, \email{nils.wisiol@tu-berlin.de }}

%
\maketitle              

\begin{abstract}
The SSH protocol is commonly used to access remote systems on the Internet, as it provides an encrypted and authenticated channel for communication.
If upon establishing a new connection, the presented server key is unknown to the client, the user is asked to verify the key fingerprint manually, which is prone to errors and often blindly trusted.
The SSH standard describes an alternative to such manual key verification: using the Domain Name System (DNS) to publish the server key information in SSHFP records.

In this paper, we conduct a large-scale Internet study to measure the prevalence of SSHFP records among DNS domain names. We scan the Tranco 1M list and over 500 million names from the certificate transparency log over the course of 26 days.
The results show that in two studied populations, about 1 in 10,000 domains has SSHFP records,
with more than half of them deployed without using DNSSEC, drastically reducing security benefits.

\end{abstract}
%
%
\section{Introduction}
The Secure Shell (SSH) protocol allows to securely establish connections to remote servers over insecure transport channels \cite{lonvick_secure_2006}.
It was standardized in Request for Comments (RFCs) 4250ff more than 15 years ago and is widely adopted on the Internet.
In May 2022, the Internet scanning service Shodan reports close to 21 million active SSH-based services found in the IPv4 space \cite{noauthor_shodan_nodate} (most of them using TCP port 22). Common use-cases include remotely administering computer systems, copying files between systems, tunneling TCP connections or graphical sessions, and many others \cite{noauthor_ssh1_nodate}.

\begin{figure}
	\centering
	\includegraphics[width=0.95\textwidth]{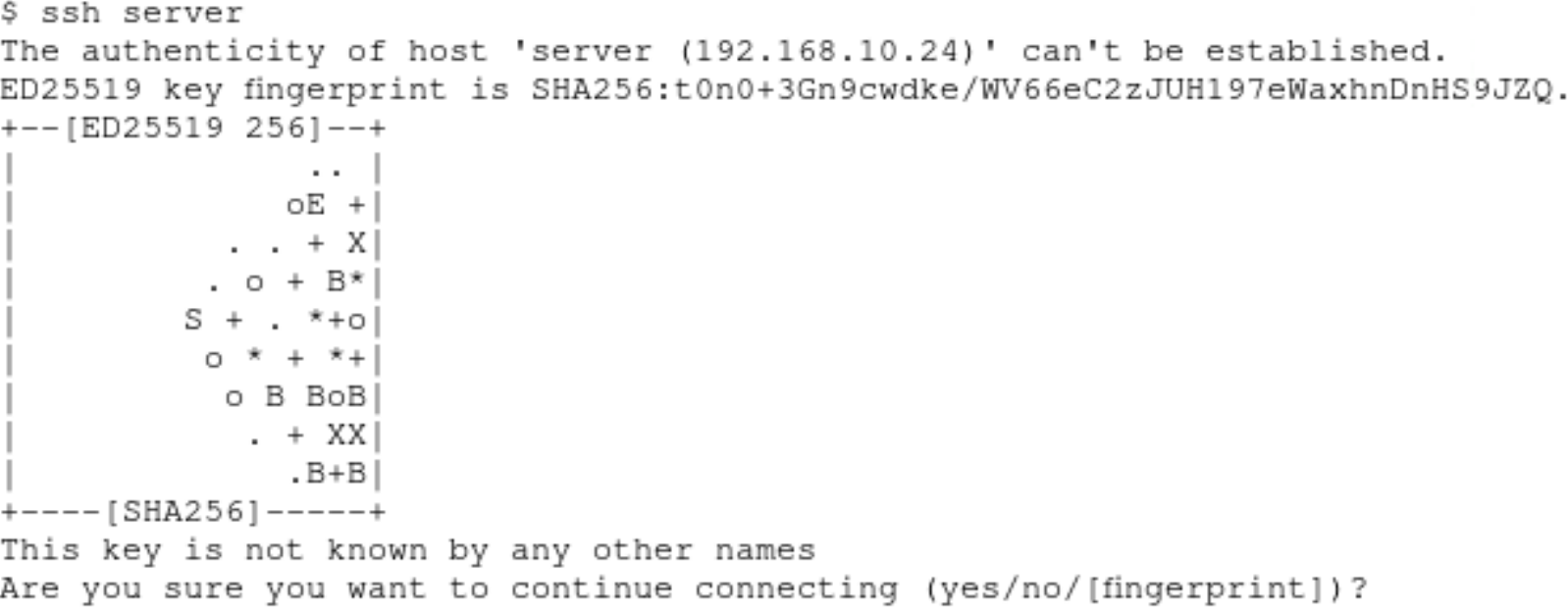}
	\caption{An example of an SSH client prompting the user to verify the SSH server's host public key fingerprint the first time a connection is established.}\label{fig:openssh-hostkeyverfication}
\end{figure}

The SSH protocol defines a server and a client component. The former offers a service, usually on a remote system, and the latter is used to connect to such services.
Upon establishing a connection, the client ought to verify the server's identity, which is done using public-key cryptography.
To verify the server's identity, the client requires the (fingerprint of the) server's public key.
If the client has no public key on record for the given host name, the key is retrieved from the server without authentication and presented to the user for manual verification, as can be seen in Fig. \ref{fig:openssh-hostkeyverfication}.
Only if the user chooses to trust the key, a connection is established.
The client then stores the server's host name and fingerprint information and uses it to verify host keys presented in future connection attempts.
This Trust On First Use (TOFU) principle makes this initial interaction pivotal to the security of all SSH connections from this client to this host name. An adequate validation requires the user to obtain the host key fingerprint through an out-of-band channel, i.e., asking the server's administrator. However, using these channels is not always feasible or comfortable. Thus users might accept fraudulent host keys or unknowingly perform human errors while comparing the fingerprints, as anecdotal evidence shows according to \cite{gutmann_users_nodate}.

The risks associated with blindly trusting SSH host keys are many-fold. For example, if password-based authentication is used, a malicious SSH server can collect and store the user's password in plain text using publicly available tools \cite{oosterhof_cowrie_2022}. Further, the OpenSSH client sends all of the client's public keys to the server, which can be used to personally identify the user through correlation with other sources \cite{valsorda_whoamifilippoio_2022}.
Although alternative, password-less authentication methods mitigate the credentials-stealing attacks,
in an online scenario, attackers in an unauthenticated connection can still use the user's credentials to obtain access to the SSH service under attack and deposit the attacker's key as trusted into the client's key database.

One approach to avoid manual verification of SSH server keys is to store the server's host key fingerprints in the Domain Name System (DNS) and let the SSH client use this out-of-band channel to retrieve them. RFCs 4251 and 4255 described the use of such SSHFP records in detail \cite{lonvick_secure_2006,griffin_using_2006}.
With DNSSEC seeing more widespread adoption \cite{noauthor_why_2020,chung_longitudinal_2017}, information retrieved via the DNS can be verified to be authentic as provided by the entity that owns the host name of SSH server.
Consequently, the SSHFP specification mandates that the information obtained from the DNS can only be trusted if DNSSEC is used \cite{griffin_using_2006}.
Hence, to enable host key verification based on SSHFP, DNS name owners are responsible for setting up the correct SSHFP records and securing their zone using DNSSEC.

\subsection{Related work}
Different aspects of SSH were covered by academic work in the past. The widely used OpenSSH server and client implementation were studied with respect to their security. \cite{albrecht_plaintext_2009} presents a variety of plaintext-recovering attacks against SSH, while \cite{miller_security_nodate} reports on the security measures to make OpenSSH more resilient against vulnerabilities. The vulnerability database \emph{CVE details} lists less than 100 discovered security issues in over 20 years to this date \cite{noauthor_openbsd_nodate}. Since OpenSSH bases its cryptographic guarantees on underlying libraries, such as OpenSSL, it is affected by issues in those. An extensively studied incident is a bug in OpenSSL which lead to predictable random numbers endangering SSL/TLS and SSH servers \cite{ahmad_two_2008,yilek_when_2009,heninger_mining_nodate}.
A recent study analyzed OpenSSH's update patterns and discovered that many servers lag behind current patches, rendering them vulnerable \cite{west_longitudinal_2022}. Further, privacy implications of SSH authentication mechanisms were studied by \cite{kannisto_time_2017} and \cite{valsorda_whoamifilippoio_2022}.

While the previously mentioned work shows that SSH is being studied intensively, there is little research on the crucial first step of using SSH: The first connection to a server and its host key verification. As outlined in the introduction, improper handling of the initial verification puts users at risk. To the best of our knowledge, only \cite{gasser_deeper_2014} performed a large-scale analysis of SSH servers and SSHFP records. They report that only 660 of 2,070 domains with SSHFP records can be authenticated using DNSSEC and thus highlight the importance of correctly securing these records. Although \cite{krishnaswamy_dnssec_2009,hardaker_enabling_nodate} and \cite{buijsman_automatic_nodate} have contributed work to facilitate DNS-based host key verification, this method appears to be lacking adoption or proper security configuration.

\subsection{Contributions}
In this paper, we set out to analyze the current situation of DNS-based host key verification information on a large scale. In essence, we answer the following research questions:
\begin{enumerate}
	\item How common are DNS-based host key verification records (SSHFP)?
	\item Do the SSHFP records match their service counterpart?
	\item Are these records properly secured using DNSSEC?
\end{enumerate}
We answer these questions with large-scale, in-the-wild measurements with respect to the prevalence, correctness, and authenticity of SSHFP resource records according to the respective RFCs based on the Tranco Top 1M domains as well as more than 500 million domains obtained from the certificate transparency log over the course of 26 days, complementing and updating prior work.

\newpage
\section{Background}
\subsection{Host Key Verification}
\subsubsection{Server-side keys}
A set of key pairs is required to run an SSH server. The keys are used for authentication of the server against the connecting user.
If no keys are given, OpenSSH generates several key pairs using different algorithms (ECDSA, ED25519, and RSA) on the first start-up \cite{noauthor_sshd8_nodate}.
Previous versions of OpenSSH also featured DSA key pairs \cite{noauthor_sshd8_nodate-1}.

\subsubsection{Fingerprints}
The host key verification process requires validating if the server's public key matches the expected one. This check is performed each time a connection to the server is being established.
To ease the process for the user, in OpenSSH, the verification of the host key fingerprint is facilitated using a hashed and encoded format using SHA256 and base64.

\subsubsection{Client-side verification}
When connecting to an SSH server, the server's public host key is sent to the client, which then tries to verify it \cite{noauthor_sshd8_nodate}. There are three possible outcomes:
\begin{enumerate}
	\item The host key is known and trusted.
	\item The host key is unknown, and no entry for the host name exists in the local database.
	\item The host key is unknown, but there is an entry for the host name in the local database.
\end{enumerate}
In case 1, the user connects to a server whose host key is already trusted and stored in its local database. Both, the host names and key information match, and the connection succeeds.

If the server key cannot be found in the local database (case 2), the user is asked to verify the server's host key. This step is prone to human error, i.e., misreading the fingerprint, ignoring it due to inconvenience or other factors \cite{gutmann_users_nodate}. However, it is crucial for the TOFU principle to confirm that the fingerprint corresponds to the server's identity. Listing \ref{lst:server-tofu-connection} displays the OpenSSH's prompt to verify the host key. If the user accepts the host key, it will be permanently stored in its local database.
\begin{lstlisting}[caption={On the first connection, the user is informed that the authenticity cannot be established and asked to verify the host key fingerprint.}, label={lst:server-tofu-connection}]
	The authenticity of host 'server (192.168.10.24)' can't be established.
	ECDSA key fingerprint is SHA256:jq3V6ES34fNDKdn5L1sbmhoyJ5MN9afd9wIS1Upa1dc.
	This key is not known by any other names
	Are you sure you want to continue connecting (yes/no/[fingerprint])?
\end{lstlisting}
When a connection to a known host name is established, but the presented server host keys do not match the record on file, either due to administrative changes on the server or an attacker performing a man-in-the-middle attack, the host key fingerprint comparison will fail due to a mismatch. In this case (3), the OpenSSH client implementation displays an eye-catching warning as seen in Listing \ref{lst:server-changed-hostkey}. The user is made aware of that issue, and the connection is aborted. To resolve the issue, the user can either remove the known fingerprint and re-accept the key or add the correct fingerprint to their database (leading to prior scenarios).

In the previously discussed cases 2 and 3, the user needs to obtain the host key fingerprints for the server using an authenticated out-of-band channel. The RFC 4251 (\cite{lonvick_secure_2006}) suggests obtaining the key via telephone or the DNS. To avoid human error and in order to enable cryptographic authentication of the obtained fingerprint, RFC 4255 \cite{griffin_using_2006} standardizes the key exchange using DNS with security extensions. This functionality is included in the OpenSSH client, but not enabled by default \cite{noauthor_ssh_config5_nodate}.
If enabled, OpenSSH attempts to retrieve fingerprints for unknown host names via DNS, and if a successful match is produced, the user is asked to continue connecting without a manual fingerprint verification.
\begin{lstlisting}[caption={A large warning is displayed and the connection aborted when the host key fingerprints do not match.}, label={lst:server-changed-hostkey}]
@@@@@@@@@@@@@@@@@@@@@@@@@@@@@@@@@@@@@@@@@@@@@@@@@@@@@@@
@  WARNING: REMOTE HOST IDENTIFICATION HAS CHANGED!   @
@@@@@@@@@@@@@@@@@@@@@@@@@@@@@@@@@@@@@@@@@@@@@@@@@@@@@@@
IT IS POSSIBLE THAT SOMEONE IS DOING SOMETHING NASTY!
Someone could be eavesdropping on you right now (man-in-the-middle attack)!
It is also possible that a host key has just been changed.
The fingerprint for the ECDSA key sent by the remote host is
SHA256:jq3V6ES34fNDKdn5L1sbmhoyJ5MN9afd9wIS1Upa1dc.
Please contact your system administrator.
Add correct host key in  ~/.ssh/known_hosts to get rid of this message.
Offending ED25519 key in ~/.ssh/known_hosts:121
Password authentication is disabled to avoid man-in-the-middle attacks.
Keyboard-interactive authentication is disabled to avoid man-in-the-middle attacks.
UpdateHostkeys is disabled because the host key is not trusted.
server: Permission denied (publickey,password).
\end{lstlisting}

\subsection{SSHFP records}\label{sec:sshfp-records}
\subsubsection{Format}
Publication of SSH server key fingerprint in the DNS is done using the \emph{SSHFP} resource records type and is documented in RFC 4255 \cite{griffin_using_2006}. SSHFP records include three fields;  Listing \ref{lst:sshfp-record-structure} shows the structure.
\begin{lstlisting}[caption={SSHFP record presentation format \cite{griffin_using_2006}.}, label={lst:sshfp-record-structure},breaklines=false]
	SSHFP <KEY-ALGO> <HASH-TYPE> <FINGERPRINT>
\end{lstlisting}
The first field, \texttt{KEY-ALGO}, is an identifier for the host key's cryptographic algorithm. The second field, \texttt{HASH-TYPE}, encodes the hashing algorithm used to generate the fingerprint as an integer. Finally, the third field, \texttt{FINGERPRINT}, contains the actual fingerprint.
Tab. \ref{tab:sshfp-key-types} and \ref{tab:sshfp-hash-algos} provide an exhaustive mapping of valid \texttt{KEY-ALGO} and \texttt{HASH-TYPE} values, according to the Internet standards \cite{griffin_using_2006,sury_use_2012,moonesamy_using_2015,harris_ed25519_2020}.

\begin{table}
	\centering
	\parbox{.45\linewidth}{
	\centering
	\caption{Values for the SSHFP \texttt{KEY-ALGO} field.}\label{tab:sshfp-key-types}
	\begin{tabular}{lll}
		\toprule
		{\bfseries Value} &  {\bfseries Algorithm} & {\bfseries RFC}\\
		\midrule
		0 & reserved & 4255\\
		1 & RSA & 4255\\
		2 & DSA & 4255\\
		3 & ECDSA & 6594\\
		4 & ED25519 & 7479\\
		5 & unassigned\cite{noauthor_dns_nodate} & - \\
		6 & ED448 & 8709\\
		\bottomrule
	\end{tabular}
	}
	\hfil
	\parbox{.45\linewidth}{
	\centering
	\caption{Values for the SSHFP \texttt{HASH-TYPE} field.}\label{tab:sshfp-hash-algos}
	\begin{tabular}{lll}
		\toprule
		{\bfseries Value} &  {\bfseries Algorithm} & {\bfseries RFC}\\
		\midrule
		0 & reserved & 4255\\
		1 & SHA1 & 4255\\
		2 & SHA256 & 6594\\
		\bottomrule
	\end{tabular}
	}
\end{table}

\subsubsection{Matching}
Several conditions must be met when matching SSHFP records with server-side host keys. Unless all the following conditions are fulfilled, the connection is vulnerable to man-in-the-middle attacks \cite{griffin_using_2006}: (1) The key algorithms must match. (2) The key's fingerprints, calculated using the designated hashing algorithm, must match. (3) The SSHFP records must be trustworthy, i.e., received and validated using DNS with DNSSEC or another secure transport channel.

\subsubsection{DNSSEC}
In light of additional administrative efforts, the adoption of DNSSEC is only progressing very slowly \cite{wander_measurement_2017,decker_reproduction_2020}. If records have been authenticated, this is indicated by the AD bit in the DNS response, which assumes that the network between client and DNS resolver is trusted.
One goal of this paper is to establish whether SSHFP records are correctly configured and secured.

\section{Methodology}

\subsection{Data collection}
\begin{figure}
	\includegraphics[width=\textwidth]{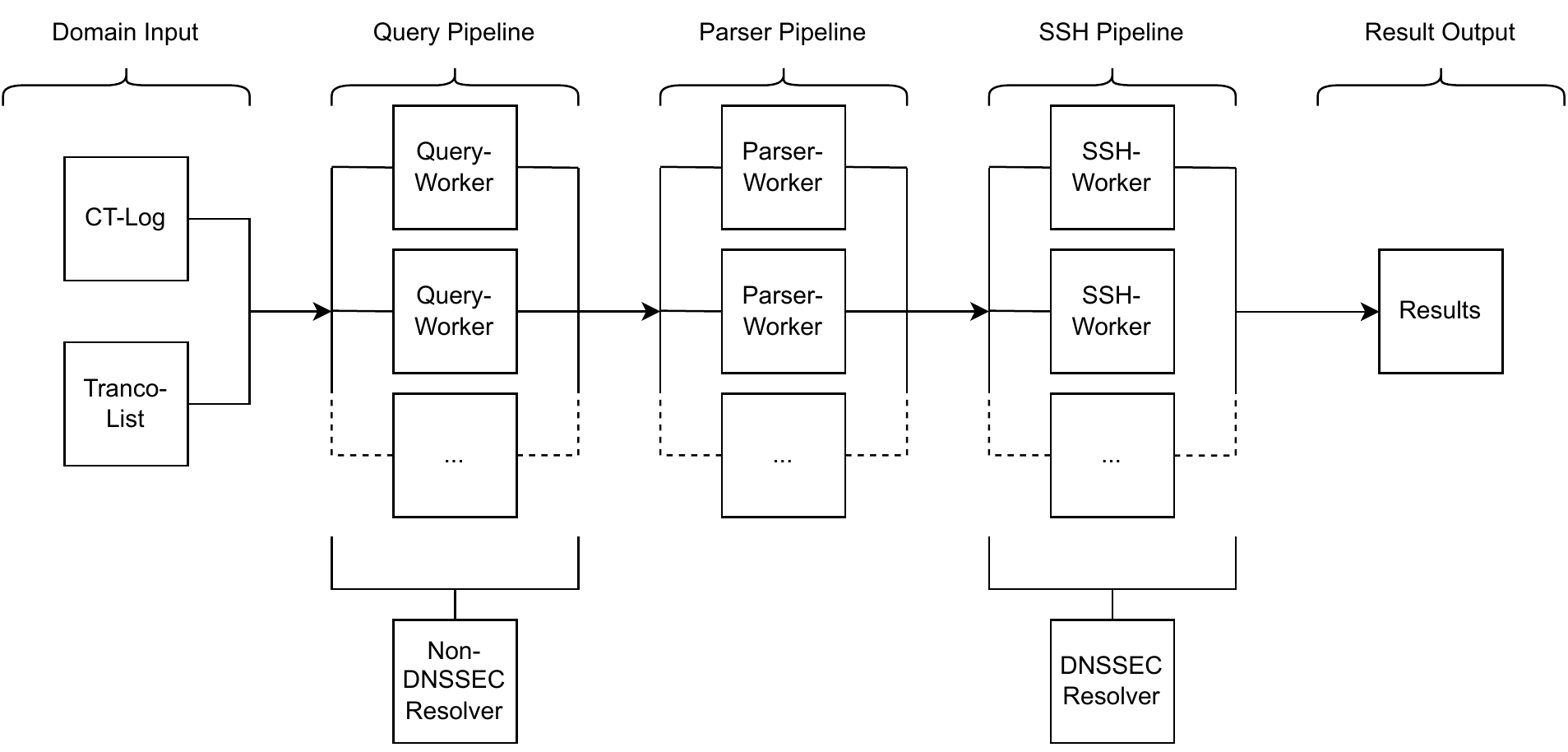}
	\caption{The data collection tool consists of multiple pipelines used for scanning and analyzing the domains for SSHFP records.}\label{fig:scanning-pipeline}
\end{figure}

To collect bulk data from the DNS and SSH servers, we implemented a scanner using Python and multiprocessing that can be easily extended for other collection tasks and is available on GitHub under a free license \cite{ourcode}. Fig. \ref{fig:scanning-pipeline} shows the stages of our methodology.

\subsubsection{Domain Input}
For our study, we relied on two sources for domain names: First, the Tranco list \cite{le_pochat_tranco_2019}, which includes 1 million popular domain names (as of 6 Dec 2021, ID: G8KK). Second, names published in the certificate transparency log \cite{laurie_certificate_2013}, which includes all names for which certificates have been issued over the course of 26 days (22 Dec 2021 to 18 Jan 2022). Together, we scanned about 515 million domain names for the presence of SSHFP records.

This methodology based on domain names complements prior work (\cite{gasser_deeper_2014}), which studied the prevalence of SSHFP records in the entire IPv4 Internet, but relied on reverse lookups, i.e., names associated with an IPv4 address. The limitation of this methodology is that not for all addresses of SSH servers, the reverse lookup pointing to a domain name is correctly configured. The limitation of the methodology used in this work, in contrast, is that not all SSH server addresses appear in our sample of domain names.

\subsubsection{Query Pipeline}
The Query Pipeline workers query a domain for SSHFP records using a shared PowerDNS\footnote{\url{https://www.powerdns.com/}} recursive DNS resolver. For any given domain name, this step returns zero or more SSHFP records. If records are present, the data is passed on to the next stage. Otherwise, the domain is omitted from further processing. Our scanning configuration consisted of 50 Query workers.
\subsubsection{Parser Pipeline}
The third stage parses a domain's SSHFP records and checks them for syntactical correctness, i.e., if a record adheres to the format described in Sec. \ref{sec:sshfp-records} and if the fingerprint length matches the output length of the hash function specified in the given \texttt{HASH-TYPE}. Erroneous and syntactical correct records are logged; domains with syntactically correct records are passed to the SSH Pipeline.
\subsubsection{SSH Pipeline}
For each given domain name, all associated Internet addresses are determined by retrieving the A record sets for the given name. For each IPv4 address, the SSH fingerprints are collected using \texttt{ssh-keyscan}\footnote{\texttt{ssh-keyscan -D -4 -t dsa,rsa,ecdsa,ed25519 -T 5 <IP>}}. (To avoid expensive port scans, we only consider SSH services on TCP port 22.) Unless the set of these fingerprints is empty, i.e., no SSH Server was reachable, they are saved for later comparison. Additionally, the SSHFP query is repeated, but this time through the DNSSEC-validating resolver (cf. Fig. \ref{fig:scanning-pipeline}) to determine whether these records can be authenticated. Again, the results, including the failure-cases, are logged to a file for later analysis.
We did not contact IP v6 SSH services due to a lack of support in our measurement network. Of 17,759 domain names with SSHFP records studied in this work, 8,200 had IPv4 and IPv6 addresses assigned and are thus only partially included (but note that it is common for dual-stack setups that v4 and v6 address point to the same service); 182 domain names (1\%) are fully excluded as they only had v6 addresses.

All SSH connection attempts were made from a single host in our institution's network, using a single IP address.
To avoid being blocked by firewalls and intrusion detection systems, no login attempts were made, and the connection was limited to obtaining the SSH key fingerprints.

\subsubsection{Result Output}
During the previous pipeline step, the collected information is logged as JSON-formatted lines by the logging framework in separate files. This data is later used as the basis for evaluating the scanning results and is publicly available \cite{ourdata}.
\subsubsection{DNS Resolvers}
To avoid problems with direct upstream resolvers, we ran two instances of the freely available PowerDNS resolvers. Apart from DNSSEC, default settings were used. The primary instance was configured without DNSSEC validation to reduce the load on the system and network. The second instance featured DNSSEC validation to establish the SSHFP records' security, but the workload was assumed to be much lower due to the previous filtering steps.

\subsubsection{Ethical considerations}
We minimized the risk and harm to the Internet caused by our research by taking the following steps.

First, we scanned from our university's network using a single IP address to make the network traffic easily recognizable as research activity. Further, we set up a website stating our intentions and research activities on that particular IP address. It included a point of contact to have all scanning activities ceased and have the domain removed from our lists.

Second, we used locally hosted recursive DNS resolvers to distribute the query-load on the respective upstream name servers. Our query-per-second rate averaged in the low hundreds, which we assume not to cause any harm to any network.

Third, SSH connection attempts were only performed if several prerequisites were met, i.e., an SSHFP and A record were found, thereby significantly reducing the number of connections. In particular, our SSH connections only query the server's host keys and do not attempt a login that might trigger various intrusion detection systems.

\subsection{Data analysis}

\subsubsection{Empirical data and quantitative analysis}
As stated previously, we empirically collected data to analyze the prevalence of SSHFP records on the Internet. Since the analysis of the Tranco 1M domains only showed an extremely small prevalence, we proceeded to evaluate a larger data set of domain names used in the wild, obtained from the global certificate transparency log (\cite{laurie_certificate_2013}).

We approach the resulting data sets with quantitative analysis and aim to answer our research questions with basic statistical methods. All data and analyses are available to the public (\cite{ourdata}).

\subsubsection{SSHFP library}\label{sec:sshfp-lib}
All code related to handling SSHFP records was implemented as a stand-alone Python library, which
will be published to facilitate current and future research \cite{libsshfp}. It covers all basic functionality to work with SSHFP records and their structure. For example, records can be checked for validity, manipulated, or created. The library complements the functionality of other python DNS libraries, such as dnspython\footnote{\url{dnspython.readthedocs.io/}}.

\subsection{Replicability}
All code used in data collection and analysis for this work is available on GitHub \cite{ourcode,libsshfp} and is suitable to replicate our work. As data collection results may change over time, we also provide the data we collected \cite{ourdata}.

\section{Evaluation}

\subsection{Tranco 1M}
\subsubsection{DNS Scanning}
The scanning process finished within roughly 1.5 hours. For 953,147 out of the 1,000,000 domains, we received a DNS response without error (response status \verb|NOERROR|), of which only 105 (0.011\%) had one or more SSHFP records. In total, 465 records were collected.

\subsubsection{DNS Data Analysis}
Out of the 465 records,  2 were invalid due to specifying an incorrect value for \texttt{KEY-ALGO}, and thus removed from further processing.

As each domain can have multiple SSHFP records (i.e., for each \texttt{KEY-ALGO} or \texttt{HASH-TYPE}), the comparison of the individual record sets revealed 11 identical and 94 unique, with the most common set counted 3 times.

For single SSHFP records, we discover 422 to be unique, with 3 repetitions at most. Considering only these unique records, Tab. \ref{tab:tranco-key-algos-hash-types} displays the distribution of the key algorithms and hash types.

\subsubsection{SSH Scanning and Matching}
For 72 of the 105 domain names, we successfully obtained fingerprints from the corresponding SSH services from at least one of the given IPv4 addresses. In total, we counted 380 fingerprints received via SSH connections from 75 different IPv4 addresses.

For 66 out of 75 hosts reached under these addresses (providing a total of 256 fingerprints), we found at least one fingerprint match in the DNS data corresponding to the domain name that led us to this address. In contrast, for 9 hosts (totaling 124 fingerprints), there was no match with the fingerprint records given in the DNS.

\begin{figure}
     \begin{subfigure}[b]{0.5\textwidth}
	\centering
	\includegraphics[width=\textwidth]{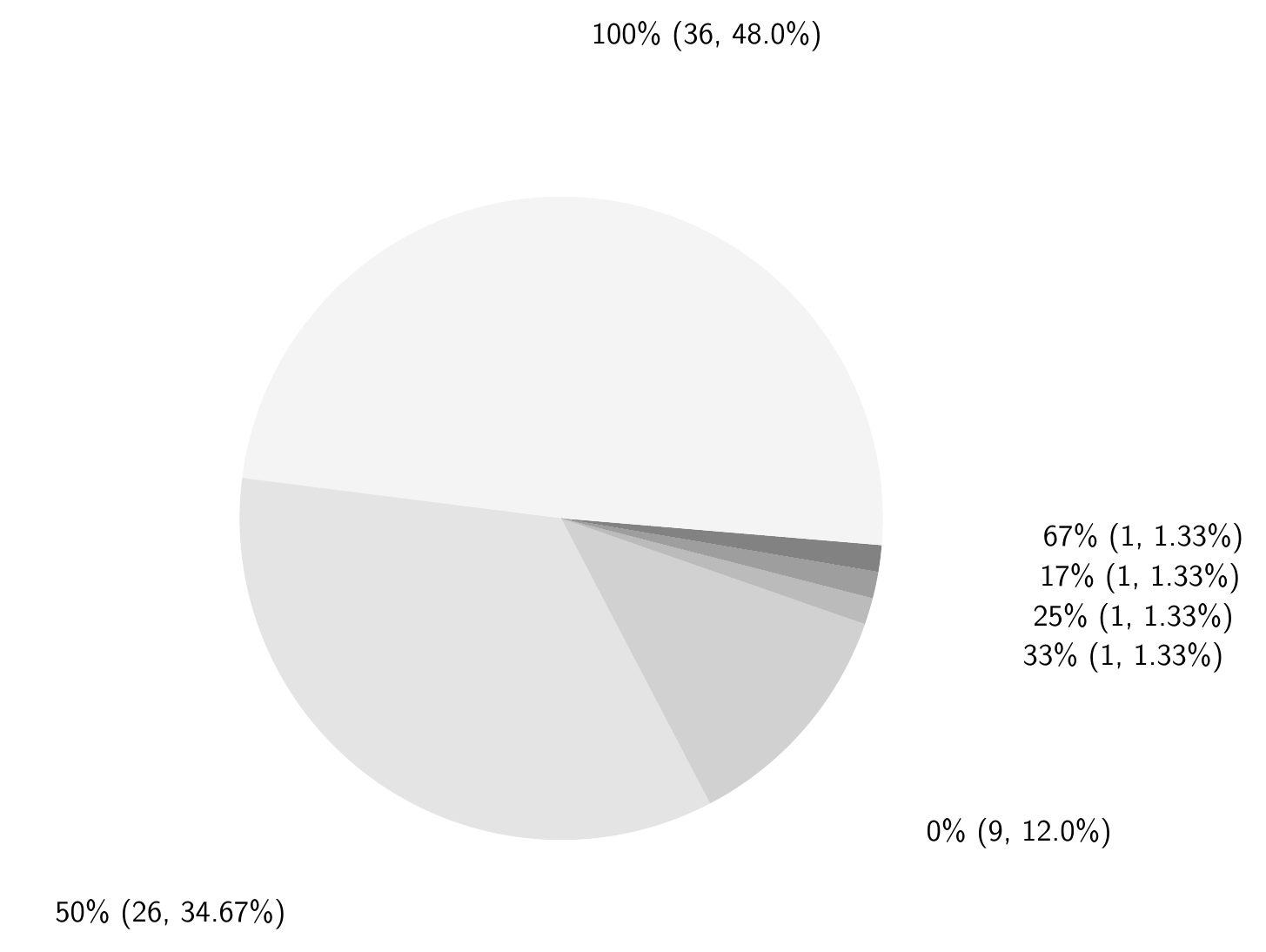}
	\caption{Tranco 1M}
	\label{fig:match-ratios-tranco}
	\end{subfigure}
	\hfill
	\begin{subfigure}[b]{0.5\textwidth}
		\centering
		\includegraphics[width=\textwidth]{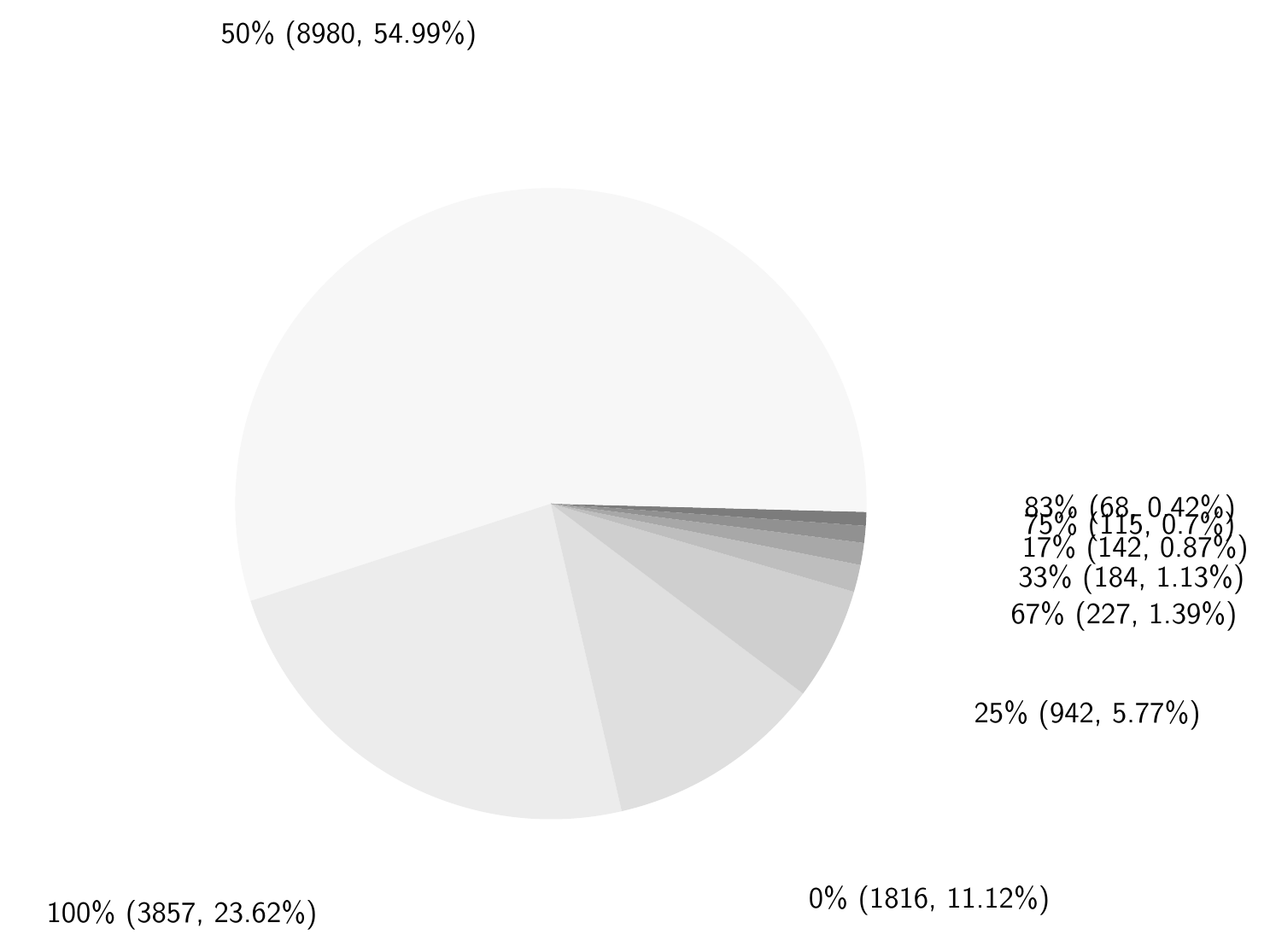}
		\caption{Certificate Transparency Log}
		\label{fig:match-ratios-certstream}
	\end{subfigure}
	\caption{Proportion of domain names that have full, partial, and no match with the fingerprint we retrieved using the SSH protocol. 100\% indicates that all fingerprints matched (implying that the same number of fingerprints was retrieved via DNS and SSH), whereas 0\% denotes no matches.}\label{fig:match-ratios}
\end{figure}

Tab. \ref{tab:tranco-key-algos-hash-types} shows the key algorithms and hash types for the matching and mismatching server-side fingerprints and Fig. \ref{fig:match-ratios-tranco} displays the distribution of Tranco domain names with full, partial, and no matching SSHFP records.

Finally, of a total of 953,147 existing domain names taken from the Tranco list, 28 (0.0029\%) had matching SSHFP records deployed securely.

\begin{table}
	\centering
	\caption{Distribution of \texttt{KEY-ALGO} and \texttt{HASH-TYPE} values for the Tranco 1M list}\label{tab:tranco-key-algos-hash-types}
	\begin{tabular}{lrrrrrrrr}
		\toprule
		{\bfseries Data From} & \multicolumn{6}{c}{\textbf{Key Algorithm}} & \multicolumn{2}{c}{\textbf{Hash Type}} \\
		&  RESERVED & RSA & DSA & ECDSA & ED25519 & ED448 & SHA1 & SHA256\\
		\midrule
		DNS & 0 & 131 & 79 & 109 & 103 & 0 & 245 & 177 \\
		SSH & 0 & 138 & 22 & 106 & 114 & 0 & 190 & 190 \\
		-- Matching & 0 & 93 & 10 & 74 & 79 & 0 & 151 & 105 \\
		-- Mismatching & 0 & 45 & 12 & 32 & 35 & 0 & 39 & 85 \\
		\bottomrule
	\end{tabular}
\end{table}

\subsection{Certificate Transparency Logs}
The evaluation for the certificate transparency (CT) log is more nuanced since the log features domains using wildcards as well as domains using many labels (i.e., \verb*|*.label2.label1.tld|).
To avoid counting duplicates, we skipped all domains using wildcards. For all other domains, to shorten domains with many labels, we identified the public suffix of each domain name. Together with the label immediately below the public suffix, the \emph{registrable domain name}, sometimes also known as eTLD+1, is derived (in above example, \texttt{label1.tld}).
If a domain occurred multiple times in the CT log, it was also scanned multiple times. On average, domain names showed up 3.8 times over the course of 26 days (median: 3). Some domains showed up frequently, and the most frequent name was seen 6 million times.

\subsubsection{DNS Scanning}
This scanning process is open-ended in its nature since the CT log continuously provides new data. Therefore, we stopped the data collection process after roughly 600 million domain names and a collection time frame of 26 days between 2021-12-22 and 2022-01-18.

After removing approximately 84.6 million wildcard domains, we issued DNS requests for the SSHFP record set for each domain name, resulting in over 515.5 million queries.
This number includes queries for the same domain name if it appears multiple times in the CT log.
In total, we queried only slightly over 136.5 million \emph{unique} domain names, which belong to about 45 million unique registrable domains.
\newpage

\subsubsection{DNS Data Analysis}
Within the 136.5 million unique non-wildcard domains, we found a total of 17,672 (0.013\%) SSHFP record sets belonging to 7,007 unique registrable domains or subdomains. Many SSHFP record sets are deployed for several names: out of 17,672 record sets, we only observed 5,961 unique sets, with the most common SSHFP record set found at 1,670 different domain names.

In the course of the 515.5 million queries for SSHFP record sets, we found a grand total of 323,655 SSHFP records (counting duplicates), of which we found 365 records to be syntactically incorrect, either because the fingerprint length did not match the output length of the stated hash type, or due to the usage of unassigned \texttt{KEY-ALGO} or \texttt{HASH-TYPE} values. The incorrect records belong to 18 unique domain names.

Tab. \ref{tab:ctlogs-key-algos-hash-types} shows the distribution of the key algorithms and hash types for the unique SSHFP records.

Due to the longitudinal data collection, some domains were analyzed multiple times, allowing us to observe changes to a domain's SSHFP records. Out measurements show 543 changes in the record sets of registrable domains. In particular, the whole record set was replaced in 539 cases. In the other 4 cases, we observed two partial removals and two partial replacements.

\begin{table}
	\centering
	\caption{Distribution of \texttt{KEY-ALGO} and \texttt{HASH-TYPE} values for the Certificate Transparency Logs}\label{tab:ctlogs-key-algos-hash-types}
	\begin{tabular}{lrrrrrrrr}
		\toprule
		{\bfseries Data From} & \multicolumn{6}{c}{\textbf{Algorithm}} & \multicolumn{2}{c}{\textbf{Hash Type}} \\
		&  RESERVED & RSA & DSA & ECDSA & ED25519 & ED448 & SHA1 & SHA256\\
		\midrule
		DNS & 1 & 7,536 & 2,367 & 6,726 & 7,191 & 2 & 9,054 & 14,769 \\
		SSH & 0 & 26,974 & 5,680 & 19,562 & 20,296 & 0 & 36,256 & 36,256 \\
		-- Matching & 0 & 15,190 & 1,528 & 11,972 & 12,211 & 0 & 21,871 & 19,030 \\
		-- Mismatching & 0 & 11,784 & 4,152 & 7,590 & 8,085 & 0 & 14,385 & 17,226\\
		\bottomrule
	\end{tabular}
\end{table}

\subsubsection{SSH Scanning and Matching}
We obtained 16,331 IPv4 addresses by querying the A records for all 17,672 domain names for which we found SSHFP records.
Contacting all addresses via SSH yielded a total of 72,512 fingerprints belonging to 11,524 unique domain names (with the remainder of domains having no reachable SSH service on port 22).

Under 14,515 addresses, an SSH service presented at least one fingerprint matching an SSHFP record published under the associated domain name (89\%). This accounts for 10,378 of 11,524 unique domain names (90\%) with SSHFP records.

Again, Tab. \ref{tab:ctlogs-key-algos-hash-types} presents the key algorithms and hash types for the matching and mismatching server-side fingerprints, and Fig. \ref{fig:match-ratios-certstream} shows the distribution of domain names with full, partial, and no matching SSHFP records.

Finally, for the certificate transparency data set, only 3,896 (0.0029\%) out of the 136.5 million domains had matching SSHFP records deployed securely.

\section{Discussion}

\subsection{Prevalence of SSHFP Records}
Our data provides an estimate for the prevalence of SSHFP records on the Internet based on the Tranco 1M list and our study of over 500 million domain names. In both cases, the rate of domain names having SSHFP records is on the order of 1 in 10,000. For the Tranco 1M list, which portrays itself ``a research oriented top sites ranking hardened against manipulation'' \cite{le_pochat_tranco_2019}, where reasonable security best practices could be expected, only 0.0105\% registrable domains have SSHFP records. For the certificate transparency data set, only 0.0014\% unique registrable domains feature SSHFP records.

While SSHFP records are not strictly required for each domain, our expectation of a more widespread adoption was not met. The records are a sensible security best practice when the domain's delegated server systems operate an SSH service. This is not the case in, for example shared web hosting environments, where direct access to the SSH servers is nonexistent or even permitted.

In prior work, Gasser et al. \cite{gasser_deeper_2014} found 2,070 domain names with SSHFP records by considering the reverse lookup information for all IPv4 addresses with active SSH hosts. From the 10,483 domain names with SSHFP records that we identified, we found that 1,931 have reverse lookup information setup for their address.
We conclude that limiting the attention to addresses with reverse lookup information excludes a large proportion of deployed SSHFP from the analysis.
Furthermore, due to the different limitations of methodology, our results do not allow a direct comparison w.r.t. the prevalence of SSHFP records to the numbers obtained by Gasser et al beyond the fact that both ours and their numbers constitute lower bounds for the total number of domain names having SSHFP record sets.

The low prevalence of SSHFP records is in spite of support for SSHFP being implemented in the commonly used OpenSSH client for nearly 20 years (although not enabled by default).
With \verb|ssh -o VerifyHostKeyDNS=ask example.com|, the client will retrieve the SSHFP records and use them for verification. Also, the openssh-package comes with a tool which outputs a server's SSHFP records to facilitate importing them into a DNS zone (\verb|ssh-keyscan -D example.com|).

\subsection{Security and Privacy}
By specification, SSH clients can only validate fingerprints obtained via SSHFP records if they can be authenticated. The only automatic approach for authentication standardized is DNSSEC.

\begin{figure}
	\centering
	\includegraphics[width=1.\textwidth]{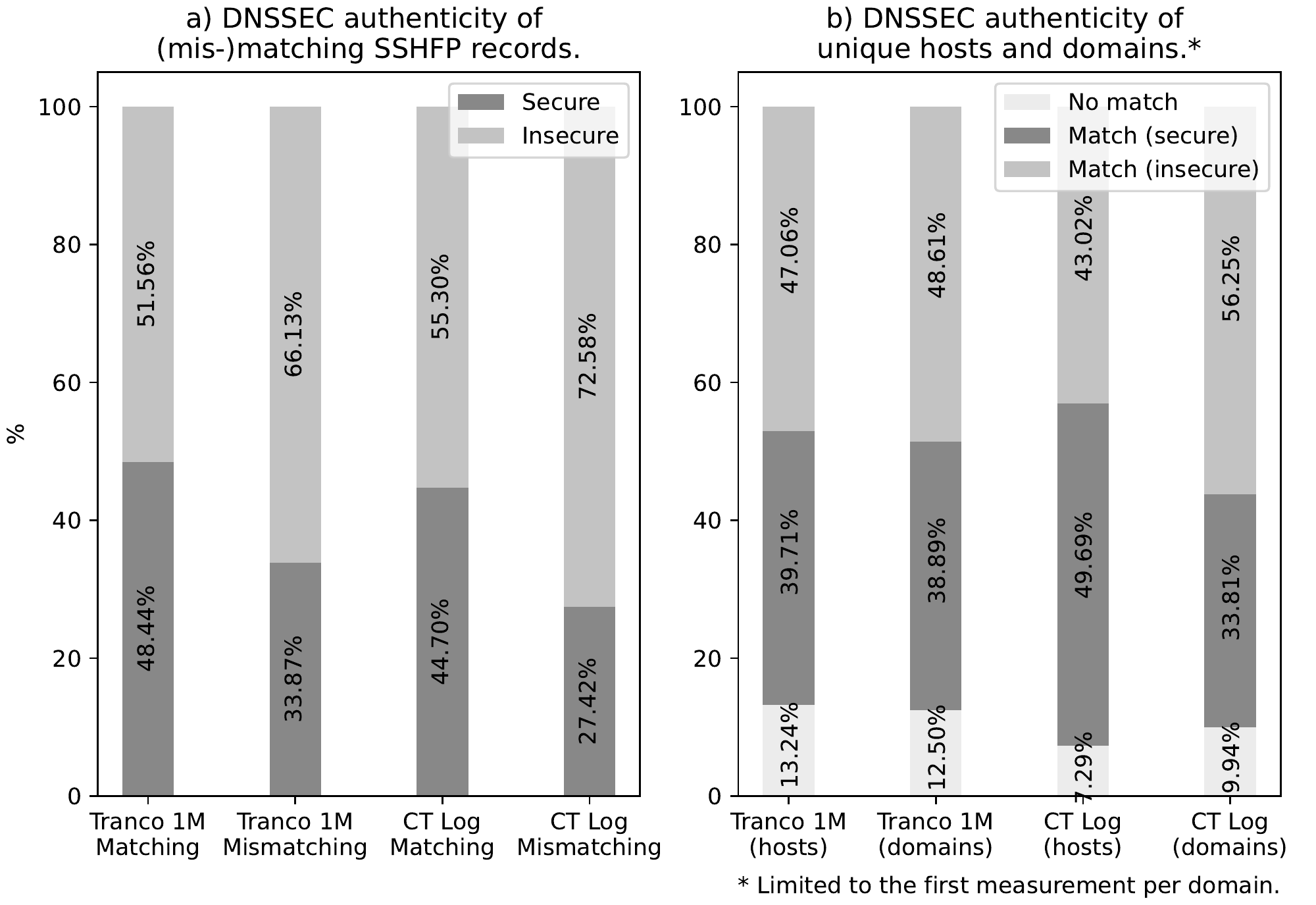}
	\caption{Overview of the DNSSEC security status of (a) SSHFP records and (b) SSHFP record sets for associated domain names and hosts (identified by IPv4 address). If multiple measurements of the same domain name are available, only the first one is considered.}\label{fig:dnssec-records}
\end{figure}

Among the Tranco 1 million domain names, out of the 72 domain names that have SSHFP records provisioned, 63 presented at least one matching fingerprint in SSH connection attempts. Of these, 28 domain names published their SSHFP records securely (i.e., with DNSSEC enabled), and 35 did so insecurely (i.e., without authenticity guaranteed through DNSSEC).
Among the 137M domain names that appeared in the CT log during our study, out of the 17,654 unique domain names that have SSHFP records provisioned, 10,378 presented at least one matching fingerprint in SSH connection attempts. Of these, 3,896 domain names published their SSHFP records securely, and 6482 did so insecurely.
Fig. \ref{fig:dnssec-records} compares the number of domains with matching fingerprints with the numbers for the Tranco domain names by DNSSEC security status.

For both populations, we find that about 88\% of SSH servers that we reached under domain names with SSHFP records actually presented matching key fingerprints. But only around half of these SSHFP records, the client can check authenticity using DNSSEC. Further, Fig. \ref{fig:match-ratios-tranco} and \ref{fig:match-ratios-certstream} show that at most around half of SSHFP records represent a full match with the fingerprints collected from the respective SSH servers. While this could be attributed to ongoing key rollovers or the precautionary deployment of emergency keys, the large prevalence of such configurations indicates that misconfigurations may be in place. The latter is supported by the observation that some mismatching records have a correct fingerprint but incorrect key algorithm or hash values.

Nonetheless, compared to \cite{gasser_deeper_2014}, we find fewer matching records (94\% vs. 88\%), which may be caused by the different methodology.
We also observe a higher adoption rate of DNSSEC (31.8\% vs. $\geq$ 44\%), which can be expected due to the overall increase in DNSSEC adoption in the 8 years since their study.

An additional insight into the used key and hash algorithms is offered by Tab. \ref{tab:tranco-key-algos-hash-types} and \ref{tab:ctlogs-key-algos-hash-types}. Modern key algorithms and hash types, such as ones based on elliptic curves and SHA-256, respectively, are still behind RSA, DSA, and SHA-1. However, the former provides higher security as, for example, the security of SHA-1 is seriously threatened \cite{stevens_first_2017}.
An interesting fact is the low number of discovered ED448 keys, despite the RFC being published more than 2 years ago, which we attribute to the default key algorithm settings of OpenSSH.

Finally, we found duplicates of several SSHFP records under various domain names, with the most popular fingerprint appearing 1,776 times.
This allows identifying related domains by observing names that share one or more host key fingerprints, which may impact privacy.
If fingerprints are identical for SSH services running on different machines, bad security practice by reusing keys is indicated. 

\subsection{Limitations}

While running this study, we noticed that a significant proportion of our initial DNS queries for SSHFP record sets of new domain names resulted in errors. Out of 516M queries, an error condition was met in 26M cases (5\%). For 16M cases (including 14k from Tranco), our resolver determined that the domain name does not exist (\texttt{NXDOMAIN}). In 10M cases, we did not receive a reply within 5 seconds (albeit retries of the resolver) or all received replies were broken.

We attribute these error conditions to faulty configuration of authoritative name servers, but caution that unreliable network links to distant servers may also play a role.
In the latter case, errors during the initial query for the SSHFP record set would incorrectly eliminate a candidate domain name from further processing; and a failed re-query at the end of the SSH-stage using the validating resolver would hinder the security assessment.
For completeness, a detailed list of such errors is available along with all data captured for this study.

Additionally, we noticed interruptions in the data provided by the certificate transparency log, due to which data in short periods of time in the study period is missing from the analysis.
The total duration of missing data is less than 3\% of the total time the study ran.

\subsection{Future Work}
A longitudinal study should be performed to better understand the evolution of SSHFP record deployment on the Internet. In addition, alternative data sources to obtain a widespread insight in the DNS or server-side host keys are an interesting area for future work,  as well as ways to a more resilient analysis process, e.g., using redundant DNS resolvers or different network locations for measurements.

Additionally, the causes of the low prevalence of SSHFP records should be studied. For example, is it an unawareness of these possibilities, the unwillingness to properly establish a server's authenticity, or technical issues (i.e., DNSSEC) which hinder the usage of SSHFP records?  Similarly, improvements for easier deployment of SSHFP records should be discussed in future work.
\newpage
\section{Summary \& Conclusion}
We provided an overview of the SSHFP specification and conducted a large-scale analysis of such records on the Internet. To that end, we collected key fingerprint information for domains from the Tranco 1M list as well as more than 500 million domains over the course of a month from certificate transparency logs. Furthermore, we obtained the corresponding key fingerprints via SSH from the addresses associated with the domain names for comparison and checked if the SSHFP records are secured by using DNSSEC.

Our data shows that SSHFP records are still a niche occurrence lacking widespread adoption, although it was mentioned early in the SSH standardization process. While a few thousand domains successfully use these records, almost 50\%  do not securely transmit them, violating the standard and drastically reducing security benefits. Nevertheless, if used correctly, SSHFP records have the potential to mitigate many risks associated with SSH's TOFU approach.

The OpenSSH software package already supports the generation and verification of SSHFP records. We complement this by publishing all code used in this work to facilitate future work in this area.

\newpage
\bibliographystyle{splncs04}
\bibliography{SSHFP}
\end{document}